\documentclass[12pt]{article}
\usepackage{cite}
\begin{document}

\title{\bf Minimal length, maximal momentum and the entropic force law}

\author{    Kourosh Nozari$^{a,}$\thanks{knozari@umz.ac.ir}\,,
            Pouria Pedram$^{b,}$\thanks{p.pedram@sbu.ac.ir}\,, and
            M. Molkara$^{c,}$\thanks{molkara.marziye@gmail.com}\vspace{.5cm}\\
            $^{a}${\small Department of Physics, Faculty of Basic Sciences,}\\
            {\small University of Mazandaran, P. O. Box 47416-95447, Babolsar, Iran}\vspace{.25cm}\\
            $^{b}${\small Department of Physics, Science and Research Branch, Islamic Azad University, Tehran, Iran}\vspace{.25cm}\\
            $^{c}${\small Department of Physics, Central Tehran Branch, Islamic Azad University, Tehran, Iran}
        }

\maketitle
\begin{abstract}
Different candidates of quantum gravity proposal such as string
theory, noncommutative geometry, loop quantum gravity and doubly
special relativity, all predict the existence of a minimum
observable length and/or a maximal momentum which modify the
standard Heisenberg uncertainty principle. In this paper, we study
the effects of minimal length and maximal momentum on the
entropic force law formulated recently by E. Verlinde.
\end{abstract}

\section{Introduction}
Various approaches to quantum gravity proposal support the idea that
near the Planck scale, the standard Heisenberg uncertainty principle
should be replaced by the so-called Generalized Uncertainty
Principle (GUP) (see \cite{1,2} and references therein). In fact,
string theory, loop quantum gravity, non-commutative geometry,
doubly special relativity and the TeV black hole physics all
indicate the existence of a minimum observable length and/or a
maximal momentum\footnote{Note that we use the phrase \emph{maximal
momentum} instead of \emph{maximum measurable momentum}. This is
because while with lower bound for position fluctuations, one can
rightfully claim that there is a minimum \emph{measurable} distance,
the way from an upper bound of momentum fluctuations to a maximum
measurable momentum is not so clear. In fact, existence of an upper
bound for momentum fluctuations just means that momentum
measurements cannot be arbitrarily imprecise, but it says nothing
about the measured momentum values (or momentum expectation
values).} in the high energy, quantum gravity era \cite{8,4,5}. The
existence of these natural cutoffs has phenomenologically
interesting implications in all high energy physics problems (see
for instance \cite{3,6,7,p} and references therein). Although the
order of magnitude of the corresponding quantum gravity corrections
are generally very small, these are direct footprints of the Planck
scale physics and contain some key attributes of ultimate quantum
gravity scenario. The idea that a gravitational system could be
regarded as a thermodynamical system has attracted a lot of
attention recently (see \cite{9} and references therein). In this
viewpoint, E.~Verlinde  has reported his achievement on the issue of
Entropic Force \cite{10}. Verlinde suggested that gravity has an
entropic origin. He postulated that the change of entropy near the
holographic screen is linear in the displacement $\Delta x$, namely
\begin{eqnarray}\label{eq1}
\Delta S=2\pi k_B \frac{mc}{\hbar}\Delta x
\end{eqnarray}
where  $m$  is the mass of the  test particle. The effective
entropic force acting on the test particle due to the change in
entropy obeys the first law of thermodynamics
\begin{eqnarray}\label{eq2}
\Delta W=T\Delta S=F\Delta x,
\end{eqnarray}
where $T$  is the temperature of the holographic screen. If one
takes the Unruh temperature \cite{11} experienced by an observer in
an accelerated frame whose acceleration is  $a$ to be
\begin{eqnarray}\label{eq3}
T=\frac{1}{2\pi k_B}\frac{\hbar a}{c},
\end{eqnarray}
then, as Vancea and Santos \cite{12} have pointed out, the postulate
(\ref{eq1}) is essentially quantum in nature. They have used the
entropic postulate to determine the quantum uncertainty in the laws
of inertia and gravity.  In addition, they have considered the most
general quantum property of the matter represented by the
uncertainty principle. Vancea and Santos have postulated an
expression for the uncertainty of the entropy such that it is the
simplest quantum generalization of the postulate of the variation of
the entropy. This expression reduces to the variation of the entropy
in the absence of uncertainties.  In this way, they obtained the
following generalization of relation (\ref{eq1})
\begin{eqnarray}\label{eq4}
\Delta S=2\pi k_B\left(\frac{\Delta x}{l_c}+\frac{\Delta
p}{mc}\right)\,,
\end{eqnarray}
where $l_{c}=\frac{\hbar}{mc}$ is the Compton length. Then by using
the Heisenberg uncertainty relation
\begin{eqnarray}\label{eq5}
\Delta x \Delta p\geq \frac{\hbar}{2},
\end{eqnarray}
and replacing  $\Delta p$ in (\ref{eq4}) by $\Delta
p=\frac{\hbar}{2\Delta x}$, they obtained quantum correction of the
Newton's second law as follows
\begin{eqnarray}\label{eq6}
F(\Delta x)=ma+\frac{\hbar}{2m}\Big(\frac{\hbar
a}{c^2}-p\Big)(\Delta x)^{-2}.
\end{eqnarray}
Recently Ghosh in Ref. \cite{13} has generalized this relation  by
using the following generalized uncertainty principle
\begin{eqnarray}\label{eq7}
\Delta x_i \Delta p_i\geq \frac{\hbar}{2}\left[1+\beta\left((\Delta
p)^2+\langle p\rangle^2\right)+2\beta \left((\Delta p_i)^2+\langle
p_i\rangle^2\right)\right],
\end{eqnarray}
where $p^2=\sum_{i=1}^{3}p_ip_i$, $\beta \sim
\frac{1}{(M_{P}c)^2}=\frac{\ell_P^2}{2\hbar^2}$, $M_P=\mbox{Planck
mass}$, $M_Pc^2=\mbox{Planck energy}\approx10^{19}$GeV. Within this
framework, Ghosh has shown that the minimum length scale modifies
the entropic force law. In the presence of a minimal observable
length, the generalized uncertainty principle can be written as
\cite{1}
\begin{eqnarray}\label{eq8}
\Delta x \Delta p\geq \frac{\hbar}{2}\left(1+\beta(\Delta
p)^2\right).
\end{eqnarray}
The minimum measurable length in this case is $\Delta
x=\sqrt{\beta}\hbar$. In this way, Ghosh has found for corrected
entropic force law the following generalized expression \cite{13}
\begin{eqnarray}\label{eq9}
F=ma+\frac{\hbar}{2mc^2(\Delta x)^2}(
a\hbar-pc^2)\left(1+\frac{\beta\hbar^2}{4(\Delta x)^2}\right).
\end{eqnarray}
In what follows, we generalize this expression for the case that
there are both a minimal observable length and a maximal momentum.
In fact, the previous analysis has the shortcoming that test
particle's momentum can attain any values without upper bound. As we
have stated, based on Doubly Special Relativity theories, test
particle's momentum has also an upper bound leading to the idea of a
maximal momentum \cite{14,15}. In other words, at least the newly
proposed doubly special relativity theories impose an upper bound on
test particle's momentum and this upper bound leads to a
modification of the entropic force law.

\section{Verlinde's entropic force  law with minimal length and maximal momentum}

In the line of the mentioned progresses, here we are going to
consider the effects of the minimal observable length and maximal
momentum on the entropic force law. In other words, we add the
existence of a maximal momentum as a new ingredient to the analysis
performed by Ghosh \cite{13}. In this respect, we obtain a general
expression for entropic force law in the framework of the newly
proposed GUP by Ali {\it et al.} \cite{4,5} . To do this end, we
consider a general GUP which contains both a minimum observable
length and a maximal momentum. The new ingredient of this GUP is
realized through those terms that are related to the existence of
$\langle p\rangle^2$ in equation (\ref{eq7}). In this respect, we
start with equation (\ref{eq7}) expressed in the following form
\begin{eqnarray}\label{eq10}
\Delta x \Delta p\geq \frac{\hbar}{2}\left[1+\beta\left((\Delta
p)^2+\langle p\rangle^2\right)\right].
\end{eqnarray}
This relation can be expressed as follows
\begin{eqnarray}\label{eq11}
\Delta p=\frac{\Delta x}{\beta \hbar}\pm \frac{\Delta x}{\beta
\hbar}\left(1-\frac{\beta \hbar^2}{(\Delta x)^2}-\frac{\beta^2
\hbar^2\langle p\rangle^2}{(\Delta x)^2}\right)^{\frac{1}{2}}.
\end{eqnarray}
To have correct limiting result when $\beta\rightarrow0$, we find
\begin{eqnarray}\label{eq12}
\Delta p=\frac{1}{2}\frac{\hbar}{\Delta
x}\left(1+\frac{1}{4}\frac{\beta \hbar^2}{(\Delta x)^2}+\beta\langle
p\rangle^2\left(1+\frac{1}{2}\frac{\beta\hbar^{2}}{(\Delta
x)^{2}}+\frac{1}{4}\frac{\beta^{2} \hbar^2}{(\Delta x)^2}\langle
p\rangle^2\right) \right).
\end{eqnarray}
From the reality of $\Delta p$, we obtain
\begin{eqnarray}\label{eq13}
\Delta x=\hbar\sqrt{\beta}\left(1+\beta\langle
p\rangle^2\right)^{\frac{1}{2}}.
\end{eqnarray}
This result can be described as a new, momentum-dependent minimal
length scale. In this framework, the entropic force law in the
presence of the minimal length, equation (\ref{eq9}), generalizes to
the following expression
\begin{eqnarray}\label{eq14}
F=ma+\frac{\hbar}{2mc^2(\Delta x)^2}(
a\hbar-pc^2)\left(1+\frac{\beta\hbar^2}{4(\Delta x)^2}+\beta\langle
p\rangle^2\left(1+\frac{1}{2}\frac{\beta\hbar^{2}}{(\Delta
x)^{2}}+\frac{1}{4}\frac{\beta^{2} \hbar^2}{(\Delta x)^2}\langle
p\rangle^2\right)\right).
\end{eqnarray}
This relation  represents  the  effects of  both minimal observable
length and  the maximal momentum as quantum gravity features  on the
entropic force acting on a quantum test particle.

From another perspective, we note that quantum commutators
originated in doubly special relativity which are consistent  also
with string theory and quantum black hole physics, ensure via the
Jacobi identity that $[x_i,x_j]=0=[p_i,p_j]$. Under specific
assumptions, these features lead to the following commutator
\cite{8,4,5}
\begin{eqnarray}\label{eq15}
[x_i,p_j]=i\hbar\left[\delta_{ij}-\alpha\left(p\delta_{ij}
+\frac{p_ip_j}{p}\right)+\alpha^2\left(p^2\delta_{ij}+3p_ip_j\right)\right],
\end{eqnarray}
where $\alpha=\frac{\alpha_0}{M_pc}=\frac{\alpha_0\ell_p}{\hbar}$.
Equation (\ref{eq15}) in 1-dimension and up to ${\cal O}(\alpha^2)$
terms gives
\begin{eqnarray}\label{eq16}
\Delta x \Delta p\geq \frac{\hbar}{2}\left[1-2\alpha\langle
p\rangle+4\alpha^2\langle p^2\rangle\right],
\end{eqnarray}
where constant $\alpha$ is related to $\beta$ through dimensional
analysis with the expression $[\beta] = [\alpha^2]$.  Like what we
have done before, we obtain
\begin{eqnarray}\label{eq17}
\Delta p=\frac{1}{8\alpha^2}\left(\frac{2\Delta
x}{\hbar}\pm\left(\frac{4(\Delta
x)^2}{\hbar^2}-16\alpha^2(1-2\alpha\langle
p\rangle)\right)^{\frac{1}{2}}\right).
\end{eqnarray}
By simplification, we get
\begin{eqnarray}\label{eq18}
\Delta p=\frac{\hbar}{2\Delta
x}\left(1+\frac{\alpha^2\hbar^2}{(\Delta x)^2}\Big(1-2\alpha\langle
p\rangle\Big)\right)\Big(1-2\alpha\langle p\rangle\Big).
\end{eqnarray}
Therefore, we obtain the following generalized entropic force law
\begin{eqnarray}\label{eq19}
F=ma+\frac{\hbar}{2mc^2(\Delta x)^2}\Big(
a\hbar-pc^2\Big)\left(1+\frac{\alpha^2\hbar^2}{(\Delta
x)^2}\Big(1-2\alpha\langle p\rangle\Big)\right)\Big(1-2\alpha\langle
p\rangle\Big).
\end{eqnarray}
The most significant difference is that in equation (\ref{eq14}) the
$\langle p\rangle$ term appears to be quadratic while in equation
(\ref{eq19}) there is linear term in momentum. We note that equation
(19) (and essentially the formalism of the present paper) shows that
the weak equivalence principle could be violated at the quantum
gravity level. In fact, as Ali has shown in Ref. \cite{17}, by
studying the Heisenberg equations of motions in the presence of GUP,
the acceleration is no longer mass-independent because of the
mass-dependence through the momentum $p$. Therefore, the equivalence
principle is dynamically violated in the GUP framework. As Ali has
stated in Ref. \cite{17}, this result agrees also with cosmological
implications of the dark sector where a long-range force acting only
between non-baryonic particles would be associated with a large
violation of the weak equivalence principle \cite{18}. The violation
of the equivalence principle has been obtained also in the context
of the string theory \cite{19}, where the extended nature of strings
are subjected to tidal forces and do not follow the spacetime
geodesics.

\section{Entropic force from GUP modified Hamiltonian}

In this section, we study the effects of the generalized uncertainty
principle on the Hamiltonians of the quantum systems in quasi-space
representation. In fact, when we consider energies beyond the Planck
energy, the usual commutator relation between the position and
momentum operators does not hold anymore. However, we can still
write the Hamiltonian in terms of such operators. In this case, some
extra terms should be added to the Hamiltonian. More precisely, the
presence of a minimal length and/or a maximal momentum adds some
terms proportional to $p^3$ and $p^4$ to the Hamiltonians of
physical systems. In the following subsections, we first consider
the case of the presence of a minimal length solely and find the
modified version of the Newton's law in two above mentioned
scenarios. Then, for the case of both a minimal length and a maximal
momentum, we obtain the relations for the uncertainty of the entropy
and their corresponding entropic forces.

\subsection{GUP with a minimal length}

Let us consider the Hamiltonian of a quantum system in the presence
of a GUP which implies a minimal length in quasi-space
representation \cite{3,6,7,p} i.e.
\begin{eqnarray}
H=\frac{p^2}{2m}+\beta\frac{\displaystyle p^4}{\displaystyle3m} +
V(x).
\end{eqnarray}

From equation (\ref{eq2}) we find
\begin{eqnarray}
\frac{p}{m}\delta p+\beta\frac{4}{3}\frac{p^3}{m}\delta p+F\delta
x=T\delta S.
\end{eqnarray}
Now we can follow two different procedures for finding the
uncertainty of the entropy. First, let us consider the Ghosh
proposal, namely \cite{13}
\begin{eqnarray}
\delta S_G=2\pi k_B \left(\frac{\delta x}{l_c}+\frac{\delta
p}{mc}\right),
\end{eqnarray}
where $l_c=\frac{\hbar}{mc}$ is the particle's Compton length. By
using $\delta p=\displaystyle\frac{\hbar}{2\delta
x}\Big(1+\frac{\beta\hbar^2}{4(\delta x)^2}\Big)$\,, (see Ref. [14])
we have
\begin{eqnarray}
F_G=ma+\frac{\hbar}{2m}\left(\frac{\hbar a}{c^2}-p-\frac{4}{3}\beta
p^3\right)\Big(1+\frac{\beta\hbar^2}{4(\delta x)^2}\Big)(\delta
x)^{-2},
\end{eqnarray}
which differs with Ghosh result \cite{13}.

The second procedure is due to Vancea and Santos (VS) \cite{12}. The
idea is writing the uncertainty of the entropy as $\delta S=2\pi k_B
\left(\frac{\delta x}{l_c}+\frac{\delta K}{mc^2}\right)$ where $K$
is the kinetic part of the Hamiltonian. For our GUP, we have
\begin{eqnarray}
\delta S_{VS}=2\pi k_B \left(\frac{\delta x}{l_c}+\frac{p\delta
p}{m^2c^2}+\frac{4}{3}\beta\frac{p^3\delta p}{m^2c^2} \right),
\end{eqnarray}
which using $\delta p=\displaystyle\frac{\hbar}{2\delta
x}\Big(1+\frac{\beta\hbar^2}{4(\delta x)^2}\Big)$ results in
\begin{eqnarray}
F_{VS}=ma+\frac{\hbar p}{2m}\left(\frac{\hbar
a}{mc^3}-1\right)\left(1+\frac{4}{3}\beta
p^2\right)\Big(1+\frac{\beta\hbar^2}{4(\delta x)^2}\Big)(\delta
x)^{-2}.
\end{eqnarray}
Comparing equations (23) and (25), we see that these two approaches
are not equivalent. In fact the difference of these forces is
proportional to the acceleration as follows
\begin{eqnarray}
F_G-F_{VS}=\frac{\hbar^2a}{2m^2c^3}\left(mc-p-\frac{4}{3}\beta
p^3\right)\Big(1+\frac{\beta\hbar^2}{4(\delta x)^2}\Big)(\delta
x)^{-2}.
\end{eqnarray}
This difference is due to different assumptions adopted for $\delta
S_{G}$ and $\delta S_{VS}$.

\subsection{GUP with a minimal length and a maximal momentum}

The modified Hamiltonian in the presence of both a minimal length
and a maximal momentum can be written as \cite{4,5,8}
\begin{eqnarray}
H=\frac{p^2}{2m}-\alpha\frac{\displaystyle p^3}{\displaystyle m}
+5\alpha^2\frac{\displaystyle p^4}{\displaystyle m}+ V(x).
\end{eqnarray}
We note that due to the presence of the cubic term of the particles'
momentum, the time reversal invariance violates as a result of the
GUP with maximal momentum. In this situation, the Hamiltonian is not
certainly the physical energy of the system under consideration.
Nevertheless, the Heisenberg equation of motion is still valid.\\

Now, from $\delta W=T\delta S$ we find
\begin{eqnarray}
\frac{p}{m}\delta p-3\alpha\frac{p^2}{m}\delta
p+20\alpha^2\frac{p^3}{m}\delta p+F\delta x=T\delta S.
\end{eqnarray}
To proceed further, we need to choose the functional form of $\delta
S$. Following the Ghosh's proposal\, $\delta S_G=2\pi k_B
\left(\frac{\delta x}{l_c}+\frac{\delta p}{mc}\right)$\,  and \,
$\delta p=\displaystyle\frac{\hbar}{2\delta
x}\Big(1+\frac{\beta\hbar^2}{4(\delta x)^2}\Big)$\,, we obtain
\begin{eqnarray}
F_G=ma+\frac{\hbar}{2m}\left(\frac{\hbar a}{c^2}-p+3\alpha
p^2-20\alpha^2 p^3\right)\Big(1+\frac{\beta\hbar^2}{4(\delta
x)^2}\Big)(\delta x)^{-2}.
\end{eqnarray}
Alternatively, using Vancea and Santos proposal for the uncertainty
of the entropy we find
\begin{eqnarray}
\delta S_{VS}=2\pi k_B \left(\frac{\delta x}{l_c}+\frac{p\delta
p}{m^2c^2}-3\alpha\frac{p^2\delta
p}{m^2c^2}+20\alpha^2\frac{p^3\delta p}{m^2c^2} \right).
\end{eqnarray}
Therefore, we obtain the corresponding entropic force where we have
both a minimal length and a maximal momentum
\begin{eqnarray}
F_{VS}=ma+\frac{\hbar p}{2m}\left(\frac{\hbar
a}{mc^3}-1\right)\left(1-3\alpha p+20\alpha^2
p^2\right)\Big(1+\frac{\beta\hbar^2}{4(\delta x)^2}\Big)(\delta
x)^{-2}.
\end{eqnarray}
For this case we have
\begin{eqnarray}
F_G-F_{VS}=\frac{\hbar^2a}{2m^2c^3}\left(mc-p+3\alpha p^2-20\alpha^2
p^3\right)\Big(1+\frac{\beta\hbar^2}{4(\delta x)^2}\Big)(\delta
x)^{-2}.
\end{eqnarray}

Again, the difference is due to different assumptions adopted for
$\delta S_{G}$ and $\delta S_{VS}$.

\section{Summary}

Gravity seems to have an entropic origin and a gravitational system
could be regarded as a thermodynamical system. This idea comes from
a thermodynamical interpretation of  gravitational field equations.
Based on the Verlinde's conjecture, the change of entropy near the
holographic screen is linear in the displacement of the test
particle about holographic screen. On the other hand, all approaches
to quantum gravity  proposal support the idea of existence of  a
minimal observable length of the order of string or Planck length.
In addition, theories such as doubly special relativity lead to
existence of an upper bound of test particle's momentum. In this
respect, these theories realize a maximal momentum too. While with
lower bound for position fluctuations, one can rightfully claim that
there is a minimum \emph{measurable} distance, existence of an upper
bound for momentum fluctuations just means that momentum
measurements cannot be arbitrarily imprecise. For this reason we
used only the phrase \emph{maximal momentum} in our setup. It is
reliable to expect that finite resolution of the spacetime points
and also an upper bound on the test particle's momentum affect the
formulation of the entropic force law. The effect of the finite
resolution of spacetime points (through existence of a minimal
observable length) on the Verlinde's entropic force law was studied
by Ghosh \cite{13}. Here we focused mainly on the simultaneous
effects of both minimal length and maximal momentum on the
formulation of the entropic force law. We generalized the entropic
force law via a phenomenological interpretation of quantum gravity
proposal which contains both a minimal observable length and a
maximal momentum. This generalization could be important in the
interpretation of entropic nature of gravity at Planck scale. We
note that the formalism presented in this paper shows that the weak
equivalence principle could be violated at the quantum gravity
level. In fact, in the presence of the GUP, the test particles'
acceleration is no longer mass-independent because of the
mass-dependence through the momentum $p$. So, the equivalence
principle is dynamically violated in the GUP framework. Finally due
to the presence of the cubic term of the particles' momentum, the
time reversal invariance violates as a result of the GUP with
maximal momentum. In this situation, the Hamiltonian is not
certainly the physical energy of the system under consideration.
Nevertheless, the Heisenberg equation of motion is still valid.

{\bf Acknowledgments}\\

The work of Kourosh Nozari is supported financially by the Research
Council of the University of Mazandaran.

\bibliographystyle{mdpi}
\makeatletter
\renewcommand\@biblabel[1]{#1. }
\makeatother

\end{document}